# Novel Giant Magnetoresistance Model using Multiple Barrier Potential

Christian Fredy Naa, Suprijadi, Sparisoma Viridi and Mitra Djamal
Physics Department, Faculty of Mathematics and Natural Science,
Intitut Teknologi Bandung, Jalan Ganesha 10, Bandung
E-mail: chris@cphys.fi.itb.ac.id

**ABSTRACT**

This paper proposes a new model for Giant Magnetoresistance (GMR) and calculates its typical graph qualitatively. The model's foundation is the microscopic mechanism in GMR, where the GMR effect can be explained by intrinsic and extrinsic potential. The potentials are spin-dependent. The potentials determine the transmission probability then it will give conductance value. Here, the multiple barrier potential model is proposed as the new GMR model. The transmission probability is determined using transfer matrix method. It was found that this model is fit qualitatively with the typical GMR graph.

## INTRODUCTION

Giant Magnetoresistance (GMR) is a device which the resistance is change by the influence of magnetic field. GMR structure is thin film composed by alternating ferromagnetic (F) and non-magnetic (N) layers. Since its discovery, the theoretical model of GMR becomes the subject of much attention (see [1] for comprehensive review). The most acceptable explanation of GMR is the spin dependent scattering. The scattering is between the electron and the potentials characteristic of the GMR materials. The scattering rates then contribute for the resistance of the GMR.

The electrons pass through the potential landscape of the GMR material. For simple F/N multilayers, the potential seen by the electrons includes the intrinsic potential of the multilayered structure and the extrinsic scattering potentials due to defects [2]. These potentials are described as the following and illustrated in Fig 1.
1. The intrinsic potential is the potential of the perfect structure. It is defined as the bulk potential of ferromagnetic and non-magnetic bulk material. The intrinsic potential also occur on the transition between ferromagnetic and non-magnetic potential, let called this potential is interface potential
2. The extrinsic potential is the potential of due to the material impurity, defect or roughness. These potentials are random which scatter around the ferromagnetic and non-magnetic material.

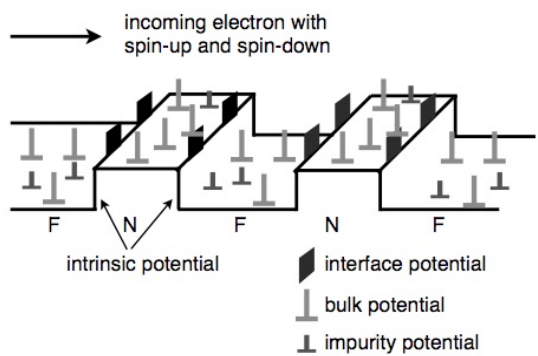

**FIGURE 1**. Potential Landscape of GMR material

Both of the potentials are spin-dependent. The bulk and interface potential that occur in ferromagnetic is due to anti-symmetry of its density of states and the transition with the non-magnetic material. The scattering potentials of impurities and defect within magnetic layers are also spin-dependent, as it is well known from experiments on bulk materials [3].

To take account all of the potentials into the GMR model is not an easy task. The Mott model [4] explained how the conductivity is measured using two different transport channels. One channel for spin-up electron and the other is for spin-down electron. This model provides qualitative result for the GMR, but still not explain the microscopic mechanism inside the GMR. However, the principle of transport channel is used in this paper.

The semi-classic model focuses on spin-dependent scattering effects. Camley and Barnas (see [5]) proposed this model and it was developed by several researcher (see [[6], [7], [8]). The major success of this model was that is predicted the thickness dependence of the GMR. However, this model neglects the influence of the intrinsic potential of the multilayer.

The first quantum mechanical model of the GMR was introduced by Levy at al. (1990) (see [9]). The model uses the Kubo formalism to calculate the conductivity of free electrons scattered by spin dependent potential. This model assumes free electrons and do not introduce the intrinsic potential of the multilayer.

With the advance in computational material science, the ab initio model (See [10], [11], [12], [13]) is introduced to the field of GMR. This model starts with first-principle calculation of the electronic structure for a perfect super lattice. The major success of this model is the calculation of band structure which can be used to determine the bulk potential. However, this model is only applicable for perfect super lattice with the absence of impurity, defect and roughness. Consequently, this model overestimates the value of MR ratio.

As a sum, to the best of our knowledge there is still no model which able to include all the intrinsic and extrinsic potential. Moreover, there is still no model that able to produce qualitatively the GMR graph (see Fig. 2). In this paper, we proposed the new model which take account of all the potentials inside the GMR and also produce the GMR graph qualitatively.

**MODEL CONSIDERATION**

All of the potentials inside the GMR material can be described using barrier potential. In this paper, one-dimensional multiple barrier potential is proposed as the new GMR model. This model is considered because of several reasons:

1. The conductivity of GMR could be pictured by the amount of transmission probability of free electrons with various energy which pass barrier potential.
2. Each potential of GMR is easily included to the model by adding the number of barrier potential. The shape of the barrier (height and width) is a function of potential kind of GMR.
3. The magnetic field dependence of the GMR is represent by the change of potential's shape either its height or its width.

In this paper, the model is built without any experimental parameters. The model is expected only to give qualitative picture of GMR as shown in Fig. 2. If the model is verified qualitatively, then the model can be used for further developments.

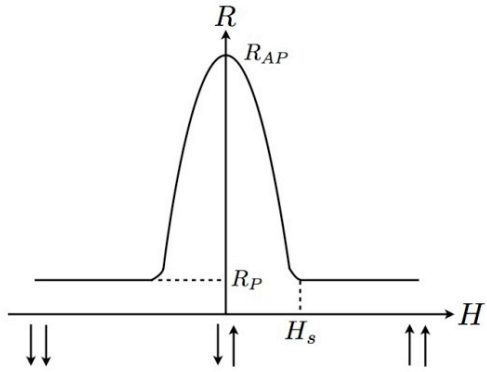

**FIGURE 2**. Typical GMR graph: Change in the resistance $R$ of the magnetic multilayer as a function of applied magnetic field $H$. The magnetization are aligned antiparallel $R_{AP}$ at zero field; the magnetization are aligned parallel $R_P$ when the external magnetic field is larger than the saturation field $H_s$.

**ONE DIMENSIONAL MULTIPLE BARRIER POTENTIAL**

Barrier potential is known as simple quantum mechanical problem where a free electron with incident wave will be transmitted and reflected by the barrier potential. This model have been used to explain the interfaces between two conducting materials and verified by the scanning tunneling microscope (STM). Important parameters of this model are the transmission and reflection coefficient. The coefficient is varied and depends on the electron's incident energy. In this model, the transmission coefficient is considered as the conductivity.

The multiple barrier potential is more complex than the single barrier. The general problem is shown by Fig. 3. One way to investigate the transport properties of the multiple barrier potential is by calculating the transfer matrix of the structure, another techniques have been given such as Green function methods, envelope function, etc. (See [14], [15], [16]). The transfer matrix is well explained by Rodriguez and Quintanilla (See [17]) to calculate the transmission probability of $Al_{0.45}Ga_{0.55}As/GaAs$ double barrier resonant tunneling (DBRT) structure.

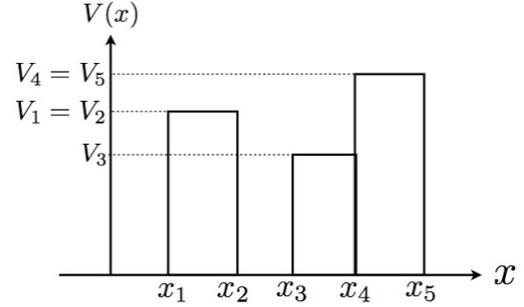

**FIGURE 3**. Multiple barrier potential problem with potential $V_n$ is associated with $x_n$.

We consider the model with the potential discontinue is occur at $x_j$ until $x_N$. The total transfer matrix for N number of discontinuities potential is formulated by:

$$M_{total} = \left[ \prod_{j+1}^{N-1} M(x_j) M(x_{j+1}, x_j) \right] M(x_n), \quad (1)$$

$$M(x_j) = \frac{1}{2} \begin{pmatrix} 1+r_j & 1-r_j \\ 1-r_j & 1+r_j \end{pmatrix}, \quad (2)$$

$$M(x_{j+1}, x_j) = \begin{pmatrix} e^{-ik_2 lx_j} & 0 \\ 0 & e^{-ik_2 lx_j} \end{pmatrix}, \quad (3)$$

where $i$ is the complex number, term $M(x_j)$ is a transition matrix between regions with different values for the potential and the term $M(x_i)M(x_{j+1},x_j)$ is a transporting matrix in a region which the potential remains constant. The term $k_j$ and $r_j$ are defined as:

$$k_j = \sqrt{\frac{2m_j}{\left(\frac{h}{2\pi}\right)^2}(E-V_j)}, \quad (4)$$

$$r_j = \frac{k_{j+1} m_j}{k_j m_{j+1}}, \quad (5)$$

where $E$ is the electron energy, $V_j$ is the potential between $x_{j+1}$ and $x_j$, $m_j$ is the effective mass, $lx_j$ is the distance between $x_{j+1}$ and $x_i$, and $h$ is the Planck constant. The transmission coefficient can be calculated using:

$$T = \frac{1}{|M_{total\,(11)}|^2}, \quad (6)$$

where $M_{total}\,(11)$ is the element (1,1) from the total transfer matrix from $x_j$ to $x_n$.

We define the conductance as the sum of transmission coefficient:

$$\sigma = \sum T(E), \quad (7)$$

and the resistance is define as:

$$R = \frac{1}{\sigma}. \quad (8)$$

**MULTIPLE BARRIER POTENTIAL FOR GMR MODELING**

Figure 4 shows the model of the multiple barrier potential. The model took two layers of non-magnetic layers (N) which sandwiched with three layers of ferromagnetic (F). The height of each barrier potential represent the bulk potential $V_{bulk}$(N) and $V_{bulk}$ (F) while the width of the potential is denotes by $\Delta w_{bulk}$ (N) and $\Delta w_{bulk}$ (F). One impurity potential $V_{defect}$ with width $\Delta w_{defect}$ is introduced in between the first two layers of ferromagnetic. The intrinsic potential is considered to be the boundary between the final layers of ferromagnetic and non-magnetic. The distances between barrier potential are denotes by $\Delta x_{bulk}$(F) and $\Delta x_{bulk}$(N). In this model, only the ferromagnetic bulk potential $V_{bulk}$ (F) which is considered to be spin dependent

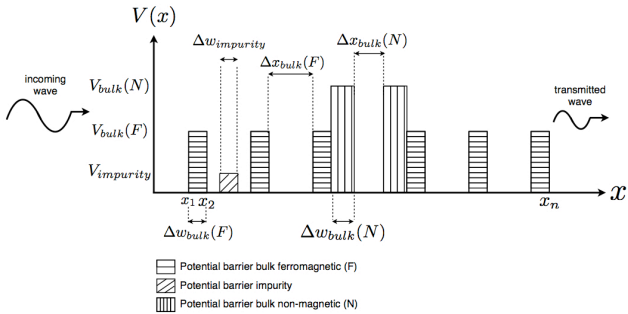

**FIGURE 4**. The multiple barrier potential model

Figure 5 shows the configuration for antiparallel (AP). Here, the incoming spin-up electrons will consider the first ferromagnetic layers as low potential (the same spin sign) while the last ferromagnetic layers as high potential (the different spin sign and vice versa with the incoming spin-down electrons. Figure 6 shows the configuration for parallel (P). Here, the incoming spin-up electrons will consider all of the ferromagnetic layers as high potential (the opposite spin sign) while the spin-down electrons will consider all of the ferromagnetic layers as low potential (the same spin sign).

This model uses dimensionless parameters. Constant $h/2\pi$ and effective mass $m$ is taken to be 1. The bulk potential for ferromagnetic layers is taken to be 2 minimum and 4 for the maximum. The bulk potential for non-magnetic or ferromagnetic layers is taken to be 2 minimum and 4 for the maximum. The bulk potential for non-magnetic layers is taken to be 1 while the defect potential is taken to be 0.5. The distance between ferromagnetic and non- magnetic bulk potential is taken to be 2, while the distance between ferromagnetic and the defect is taken to be 0.5. The electron that passes through the potentials is taken to be 3.8 until 4.8 with 200 increments.

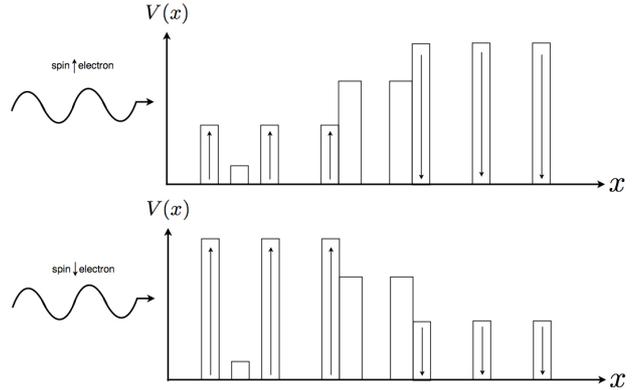

**FIGURE 5**. The antiparallel (AP) configuration

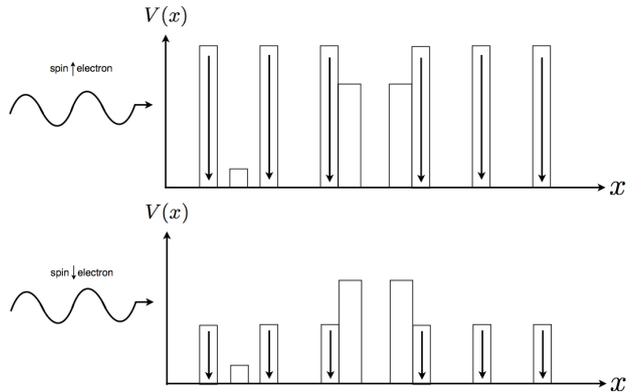

**FIGURE 6**. The parallel (P) configuration

To obtain the total resistance of the system, we use the Mott's first argument [4], the resistance is the sum of the independent resistance for the spin-up and spin-down electrons:

$$R = R_\uparrow + R_\downarrow \quad (9)$$

**RESULT AND DISCUSSION**

Figures 7, 8 show the magnetoresistance graph for normalized spin-up and spin-down channel. As shown on the figures for each channel: the electrons which have the same sign of spin obtain low resistance, while when the electrons pass the different sign of spin obtain high resistance.

The total magnetoresistance is shown in Fig. 9. This model is compared with the first generation of GMR using Fe/Cr multilayers measured by Baibich et al [18]. As shown from the comparison, the model is fit qualitatively with the experimental result. Some remarks from the model that the asymmetry is occurred because of the defect potential. Thus, it can be concluded that the defect has significant effect for the total resistance. There is also saturation condition which is also fit well with the typical GMR graph.

As a sum, it can be concluded that all of the model's result is fit qualitatively with the typical GMR graph. Thus, this model is promising for further development. In this case is to include the actual parameters from the experimental point of view.

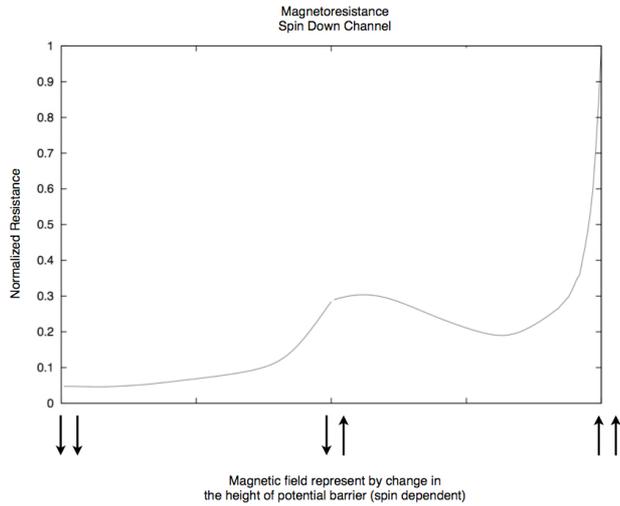

**FIGURE 7**. Spin down channel magnetoresistance graph (normalized).

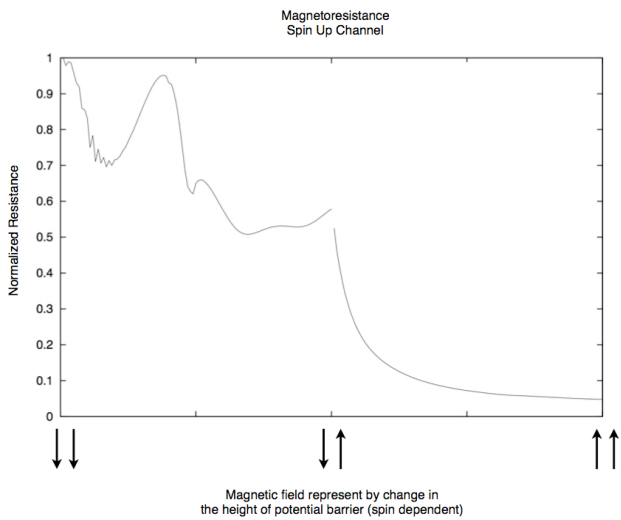

**FIGURE 8**. Spin up channel magnetoresistance graph (normalized).

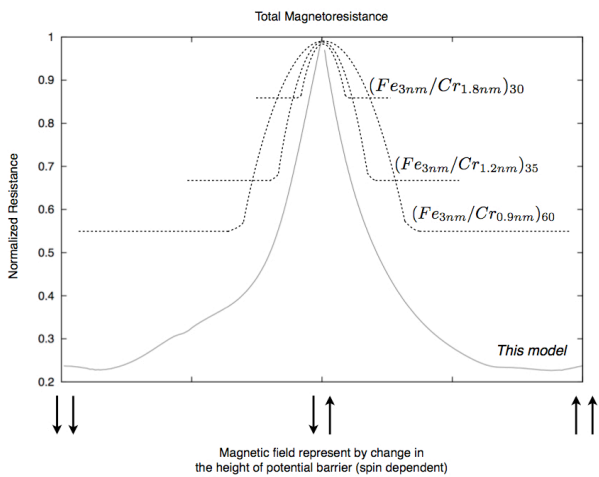

**FIGURE 9**. The total magnetoresistance (normalized) compared with variation of data of the Fe/Cr multilayer measured by Baibich et al [18].

**CONCLUSION**

This paper presents a new model for Giant magnetoresistance. This model takes account of all influenced potentials in the GMR. The model is using multiple barrier potential with its solution is determined by using transfer matrix.

In this model, the change of magnetic field is represent by the change in the height of barrier potential. The result that showed the relation between the resistance and the change on the potentials is fit qualitatively with the typical GMR graph and the experimental result.

The result shows the asymmetry graph due to the defect potential and also the saturation condition. Thus, this model is fit qualitatively with the typical GMR graph and promising for further development.

**ACKNOWLEDGMENTS**
This research is funded by I-MHERE, Faculty of Mathematics and Natural Science, Institut Teknologi Bandung.